\documentclass{article}
\usepackage{group21} \usepackage{epsf}
\title{Time of Arrival in Event Enhanced Quantum
Theory} \author{Philippe  Blanchard\adr{1} \and\
Arkadiusz Jadczyk\adr{2}} \address[1]{Faculty of
Physics and BiBoS, University of Bielefeld,
Universit\"atstr. 25, D-33615 Bielefeld, Germany;
e-mail: blanchard@physik.uni-bielefeld.de}
\address[2]{Institute of Theoretical Physics,
University of Wroc{\l}aw, Pl. Maxa Borna 9, PL-50
204 Wroc{\l}aw, Poland; e-mail:
ajad@ift.uni.wroc.pl}
\newcommand{\be}{\begin{equation}}
\newcommand{\ee}{\end{equation}} 
\begin{document}
\maketitle 
\begin{abstract}
The new solution to the problem of time of
arrival in quantum
theory is presented herein. It allows for
computer simulation of particle counters and it
implies Born's interpretation. It also suggests
new experiments that can answer the question: can
a quantum particle detect a detector without
being detected? \end{abstract}

\section{Introduction}
One of the most troublesome deficiencies of
Textbook Quantum Theory is that it leaves
questions about timing of experimental events
unanswered. The principle reason for this
deficiency is that an {\sl experimental event}
(or {\sl measurement}) can not be defined within
the standard theory \cite{bell89,bell90}. Due to
this deficit, Bell felt the need to introduce
{\sl beables} into quantum theory \cite{bell87b}.

Recently we have developed a
semi-phenomenological theory that cures this
deficiency and also has a predictive power that
is stronger than the Standard Quantum Theory.
That is why it was entitled {\sl Event Enhanced
Quantum Theory} or, in short, EEQT
\cite{blaja95ab}.

EEQT can be thought of as a formalism
implementing Bohr's idea that the end result of
any experimental event is classical in nature -
necessarily - so that we can communicate to our
colleagues what we did and what result we
obtained. The line separating the quantum from
the classical is also known as the Heisenberg
cut. EEQT is a semi-phenomenological theory
because it considers the exact placement of the
cut as a convention. Stapp (see \cite{stapp96}
and references therein) believes that the only
events that are real are mental (or {\sl
experiential}) events.  Pushing the borderline
between quantum and classical toward human
mind/brain  interface would make EEQT into a
fundamental theory - provided an exact form of
the interface between matter and mind is
identified. For most practical purposes, however,
 the borderline can be placed simply between the
quantum object and the classical measuring
apparatus or its part (e.g. its display, or
pointer). EEQT gives us the mathematical
framework to describe the interface and the
reciprocal coupling - with information flowing
from the quantum system to the classical
measuring device and with the unavoidable back
action on the quantum system.

What is most important is that EEQT provides the
algorithm that enables us to model the individual
experiential sequences, including the {\sl
timing} of events. A discussion of other apsects
of an evolutionary picture in quantum theory  has
been given by Haag \cite{haag}. Here we will
discuss its practical application in the context
of {\sl time of arrival}. \section{Time of
arrival: definition}

The simplest situation when the question of time
of arrival can be discussed is that of a particle
moving on a line and we ask at what time the
particle  will arrive at some specific point $a$
on this line. In order to answer this question
experimentally we would set a particle detector
at $a$ and measure the time interval $t$ between
the moment the particle is released and the
moment it is registered by the detector.
Experiments suggest $t$ is a random variable.
After repeating the experiment many times, and
assuming the particle is always being prepared in
the same quantum state $|\psi>$, we arrive at an
experimental probability distribution $p(t)$. We
will denote by $P(t)$ the probability that the
particle is detected up to time $t$, thus
$P(t)=\int_0^t p(s)ds$.

In practice we do not have 100\%  efficient
detectors, so we have the probability 
$P(\infty)$ that the detector will detect the
particle at all, is less than one.
\footnote{Anticipating the following discussion
let us mention at this place that numerical
simulations using our formula for time arrival
for a point--like detector suggests $P(\infty)<
0.73$}.  The standard quantum theory does not
provide us with any formula for $p(t)$.

Wigner (cf.  Eq. (5), p. 240 of Ref. 
\cite{wigner72}) has assumed, completely ad hoc,
that the formula \be p(t)= const \vert
\psi(a,t)\vert^2, \label{eq:wig} \ee, where
$\psi(t)$ is a solution of the Schr\"odinger
equation.\footnote{This formula is evidently
wrong as it leads, for a Gaussian wave packet, to
$P(\infty)=\infty$. Later on we will see that the
correct formula (cf. Eq.(\ref{eq:pst})) involves
integral transform of $\psi(a,t)$.}  One needs,
to this end, to go beyond the standard theory and
there are not so many options - one can try
Nelson's Stochastic Mechanics, Bohmian Mechanics
or EEQT.  The formula for time of arrival can
then serve as an empirical test which can judge
which of the alternatives better fits the
experimental data.

In the present paper we will follow
\cite{blaja96a} and describe the formula for time
of arrival as predicted by EEQT. In fact,
following Wigner, we will consider first a
somewhat more general problem, namely that of
time of arrival {\em at a given state} $\vert u
>$. In EEQT noiseless coupling of a quantum
system to a classical yes-no device is described
by a positive operator $F$. In our case we take
$F={\sqrt \kappa} \vert u><u\vert$, where
$\kappa$ is a phenomenological coupling constant
parameter of physical dimension $t^{-1}$. The
Master Equation describing continuous time
evolution of statistical states of the quantum
system coupled to the detector reads:
\begin{eqnarray} {\dot
\rho}_0(t)&=&-{i\over\hbar}[H_0,\rho_0(t)]+F
\rho_1 F \nonumber\\ {\dot
\rho}_1(t)&=&-{i\over\hbar}[H_1,\rho_1]-{1\over2}
\{F^2,\rho_1\}. \end{eqnarray} Suppose at $t=0$
the detector is off, that is in the state denoted
by $0$, and the particle state is $\vert \psi>$,
with $<\psi\vert\psi>=1.$ Then, according to EEQT
(cf. \cite{blaja95ab,blaja96a}) the probability
$P(t)$ of detection, that is of a change of state
of the detector, during time interval $(0,t)$ is
equal to $1-<\psi\vert K(t)^\star K(t)\vert\psi>
$, where \be K(t)=\exp (-{i\over\hbar}H
t-{F^2\over2}t). \label{eq:kt} \ee It then
follows that the probability $p(t)dt$ that the
detector will be triggered out in the time
interval (t,t+dt), provided it was not triggered
yet, is given by \be
p(t)={d\over{dt}}P(t)=\kappa\ \vert <u\vert
K(t)\vert \psi>\vert^2. \label{eq:pst} \ee The
difference between the above and Wigner's formula
is presence of the coupling constant $\kappa$ as
well as the damping term $F^2/2$ in the
definition of the propagator $K(t)$. It is this
damping term together with the coupling constant
that assure that $P(\infty) \leq 1$ in contrast
to the  formula as in \cite{wigner72}.\\ To
compute $p(t)$ let us note that $p(t)$ is equal
to $\vert \phi(t)\vert^2$, where the complex
amplitude $\phi(t)$  is given by $<u\vert
K(t)\vert \psi>$. Denoting by ${\tilde \phi(z)}$
the Laplace transform of $\phi(t)$ one easily
gets (cf \cite{blaja96a}): \be
{\tilde\phi}={{\sqrt\kappa}<u|{\tilde K_0}|\psi>
\over{1+{{\kappa\over2}\ <u|{\tilde K}_0|u>}}}
\label{eq:atp} \ee where \be K_0(t)=\exp
(-{i\over\hbar}H t). \ee This is our final
formula for the Laplace transform of the
probability amplitude of time of arrival.
\begin{figure}[ t] \epsfysize=7cm
\epsffile{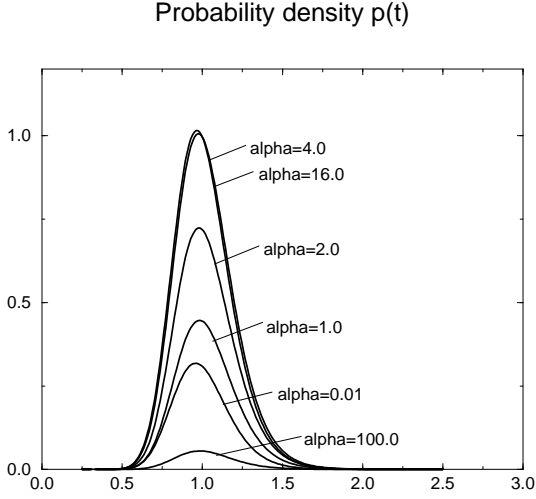} \caption{Probability
density of time of arrival for a point counter
placed at $a=0$, dimensionless coupling constant
$\alpha=m\eta\kappa/\hbar$. The incoming Gaussian
wave packet of width $\eta$ starts at $t=0$,
$x=-8\eta$, with velocity $v=2\hbar/m\eta$}
\label{fig:fig2} \end{figure}

\begin{figure}[t] \epsfysize=7cm
\epsffile{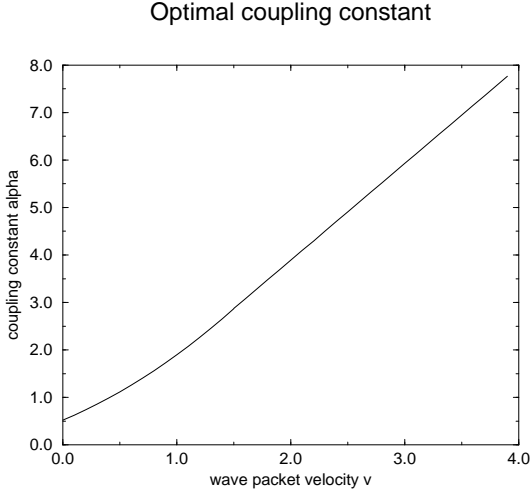} \caption{Optimal coupling
constant as a function of velocity of the
incoming wave packet. The dependence pretty soon
saturates to a linear one. At the saturation
value $P(\infty)\asymp 0.5.$} \label{fig:max}
\end{figure}

Let us consider a free Schr\"odinger particle on
a line, and let us take $u$ to denote the
improper position eigenstate at $a$, that is
$<x\vert u>=\delta(x-a)$. Then \be <u|{\tilde
K}_0|u>= {\tilde K}_0(a,a;z)=\left({\hbar
m\over{2iz}}\right)^{1\over2}. \ee Let us denote:
\be {\tilde G}(z)={1\over{1+{{\kappa\over2}\
<u|{\tilde K}_0|u>}}}= {{z^{1\over2}}\over
{z^{1\over2}+\epsilon}}, \ee where
$\epsilon={\kappa\over2}\left( {{\hbar
m}\over{2i}} \right)^{1\over2}$. It can be now
checked that the inverse Laplace transform $G(t)$
of ${\tilde G}(z)$ is given by \be
G(t)=\delta(t)+{d\over{dt}}f(t), \ee where \be
f(t)=e^{\epsilon^2t} \mbox{Erfc} \left(\epsilon
t^{1\over2}\right). \ee The amplitude $\phi(t)$
becomes then: \be
\phi(t)=\kappa^{1\over2}\left(\psi_0(t,a)+\int_0^
t{\dot f}(s)\psi_0(t-s,a) ds\right),
\label{eq:phit} \ee where $\psi_0$ stands for the
freely evolving wave function. The second term in
the formula (\ref{eq:phit}) gives the necessary
correction to the Wigner formula
(\ref{eq:wig}).\\ It is instructive to discuss
the limit of infinite coupling constant.
Numerical simulations show that for every
incoming wave packet there is an optimal value of
the coupling constant $\kappa$ which gives the
maximal efficiency of the detector. Increasing
$\kappa$ over this optimal value causes loss of
efficiency because of reflection of the particle
by the detector. In the limit of infinite
$\kappa$ the detector efficiency $P(\infty)$
drops to zero - cf. Fig \ref{fig:fig2}. One may
ask what is the maximum value of the efficiency
$P(\infty)$ for a point counter? We do not know
the answer to this question. Our guess is that
the maximum efficiency is reached for a Gaussian
wave packet that is placed centrally over the
detector with zero velocity. Numerical
simulations seems to confirm this guess. The wave
packet slowly spreads out being at the same time
"eaten" by the sink at its center. Figure
\ref{fig:max} shows the dependence of the
detector efficiency on the value of the
dimensionless coupling constant
$\alpha=m\eta\kappa/\hbar$. The maximum is
attained at the value of $\alpha\approx 1.3216$
and turns out to be $\approx 1.73$.
 It would be desirable to have an
analytical proof or disproof of our conjecture.
The fact that the maximal detector efficiency is
less than one may seem to be an artefact of the
singular character of a pointlike detector. It
may, however, also have some deeper meaning. If
so, then such a meaning is not known to us.
\section{Conclusions} We have seen that the
formula for time of arrival of a Schr\"odinger
particle contains a phenomenological parameter
$\kappa$ characterizing the strength of the
coupling between the particle and the sink. If
$\kappa$ is too small then most of the particles
would pass the detector undetected. If $\kappa$
is too big, then the sink will also act as a
reflecting barrier. For each  incoming wave
packet there is an optimal value of the coupling
that gives maximal detector efficiency.

Our formula for the time of arrival can be used
to perform again Wigner's analysis of
time--energy uncertainty relation. However, it
must be noticed that the correct analysis will be
much more difficult than that in the original
Wigner's paper \cite{wigner72}. First of all our
formula (\ref{eq:phit}) contains an extra term
which is absent in the Wigner paper. Second, in
case of a general time of arrival at a state
$|u>$ Wigner's formula for spread $\epsilon^2$
give by his Eq. (5b)\ of Ref. \cite{wigner72} is
also incorrect as his "or"  between Eq. (2) and
Eq. (2a) does not hold for a general $|u>$.

From the probabilistic point of view the process
of arrival of a quantum system at a given state
$|u>$ is an inhomogeneous Poisson process with
the rate function \be \lambda (t) = \kappa\
{{<\psi|K(t)^\star|u><u|K(t)|\psi>}\over
{<\psi|K(t)^\star K(t)|\psi>}}. \ee A more
general algorithm for a piecewise deterministic
process describing individual sample path during
a continuous measurement can be found in Ref.
\cite{blaja95ab}.

It is to be stressed that the damping term in the
propagator $K(t)$ (cf. Eq. (\ref{eq:kt}) is to be
thought of as experimentally verifiable. That is,
the very presence of a detector, even if the
particle goes through it undetected, changes the
time evolution of the wave packet by adding
imaginary potential to the Hamiltonian. The
phenomenon here is of the same kind as that
discussed by Dicke \cite{dicke81}, Elitzur et al.
\cite{eliva} and Kwiat et al. in \cite{kwiat}. We
can say that {\sl the particle can detect a
detector without being detected itself}\ . Our
formula for $K(t)$ describes this effect in a
quantitative way.

Finally, let us note that it would be interesting
to obtain a relativistic version of the time of
arrival formula.  This can be in principle done
by exploiting the ideas given in Ref.
\cite{blaja96b}.

\section*{Acknowledgments}
One of us (A.J.) would like to thank A. von
Humboldt Foundation for the support, and to L. Knight
for reading the manuscript.

\end{document}